\documentclass[11pt]{article}
\usepackage{times}
\usepackage{geometry}
\usepackage[utf8]{inputenc}
\usepackage{enumitem,amssymb}
\usepackage{graphicx}
\usepackage{fancyhdr}
\usepackage{aas_macros}
\usepackage{mdframed} 
\usepackage{hyperref}

\usepackage[authoryear]{natbib}
\setcitestyle{authoryear,open={(},close={)}}

\mdfdefinestyle{theoremstyle}{
innertopmargin=\topskip,}
\mdtheorem[style=theoremstyle]{lrptextbox}{}

\begin{document}

\begin{center}
{\huge Planetary Astronomy--\\Understanding the Origin of the Solar System} 
\end{center}

\begin{center}
{\large S.~M.~Lawler$^{1,*}$, A.~C.~Boley$^2$, M.~Connors$^3$, W.~Fraser$^4$, B.~Gladman$^2$, C.~L.~Johnson$^2$, J.J.~Kavelaars$^4$, G.~Osinski$^5$, L.~Philpott$^2$, J.~Rowe$^6$, P.~Wiegert$^5$, R.~Winslow$^{7,4}$}

1) Campion College, University of Regina, 2) University of British Columbia, 3) Athabasca University, 4) NRC-Herzberg, Victoria, 5) University of Western Ontario, 6) Bishop's University, 7) University of New Hampshire

*Samantha.Lawler@uregina.ca

\end{center}

\begin{abstract}
There is a vibrant and effective planetary science community in Canada.  We do research in the areas of meteoritics, asteroid and trans-Neptunian object orbits and compositions, and space weather, and are involved in space probe missions to study planetary surfaces and interiors.  For Canadian planetary scientists to deliver the highest scientific impact possible, we have several recommendations.  Our top recommendation is to join LSST and gain access to the full data releases by hosting a data centre, which could be done by adding to the CADC, which is already highly involved in hosting planetary data and supporting computational modelling for orbital studies.  We also support MSE, which can provide spectroscopy and thus compositional information for thousands of small bodies.  We support a Canadian-led microsatellite, POEP, which will provide small body sizes by measuring occultations.  We support the idea of piggybacking space weather instruments on other astronomical space probes to provide data for the space weather community.  Many Canadian planetary scientists are involved in space probe missions, but through haphazard and temporary arrangements like co-appointments at US institutions, so we would like the community to support Canadian researchers to participate in these large, international missions.
\end{abstract}

\section{Introduction} \label{sec:intro}

Planetary science is astronomy that can directly affect our lives here on Earth.  It’s the only part of astronomy that can support in-situ exploration: space probes with Canadian involvement are working in the Solar System right now, and there are future plans for more.  Understanding and improving space weather and NEO monitoring are important to avoid catastrophic disruption of life on Earth.  Precise measurements of many thousands of asteroids and TNOs has shaped our understanding of the formation and early dynamical history that set the planetary architecture that allowed life to evolve on Earth.  Our Solar System is one example of the outcome of planet formation and evolution that we can study in great detail;  we now know of many thousands of exoplanet systems, and we can compare our planetary architecture and small body distributions with exoplanet and debris disk systems around thousands of stars.  

For the best possible planetary science, we need continued Canadian access to optical ground-based telescopes. 
LSST will discover thousands more small bodies in the outer solar system and hundreds of thousands more in the inner solar system \citep{Ivezicetal2019}. 
Canadian involvement in LSST will be important for nearly every aspect of planetary science.  
Continued access to computation and data storage facilities in Canada as is provided by the Canadian Astronomy Data Centre (CADC) and the Canadian Advanced Network For Astronomical Research (CANFAR) will be absolutely vital in the LSST era, and provide an in-kind contribution that Canada could use to pay for LSST membership.  

In addition to LSST and associated CADC expansion, we support a small number of upcoming facilities and missions that will enable the best possible planetary science program in Canada for a low cost. The Maunakea Spectroscopic Explorer (MSE) will be able to provide spectral measurements of the compositions of thousands of small bodies. The Canadian-built microsatellite POEP will provide measurements of the size-distribution of small bodies that constrain collisional evolution models.  Setting up a process for space physicists to coordinate with astronomers in order to piggyback space weather instruments on planetary probes is an inexpensive way to increase mission science return.  Supporting Canadian scientists to participate in already existing international planetary missions is another way to greatly increase Canadian planetary science return at orders of magnitude lower cost than a Canadian-led mission.

\section{There is a vibrant and effective planetary science community in Canada}

Astronomers at institutions across Canada study a wide range of planetary science topics. Below we summarize some of the key areas Canadian astronomers are actively researching.



\subsection{Explorations of Planetary Surfaces and Interiors}

Canadian scientists are directly involved with a number of planetary science space missions. 
OSIRIS-REx\footnote{\url{https://www.asteroidmission.org}} is a spacecraft currently studying the near-Earth asteroid 101955 Bennu and will return samples in 2023. The spectral class of the asteroid suggests that it contains primitive materials from the Solar System's formation, and close examination of the asteroid can reveal the extent to which it is a rubble pile, its surface processes, and detailed information about its dynamical evolution through, for example, Yarkovsky drift. The mission has implications for planet formation, planetary defence, and in-situ resource utilization.  Canadian scientists and engineers led the development of the OSIRIS-REx Laser Altimeter \citep{daly_etal_ola}, which is producing detailed terrain maps of the asteroid's surface and is being used to select the potential locations for the sample return manoeuvre.   

InSight\footnote{\url{https://mars.nasa.gov/insight/}} landed on Elysium Planitia on the 26th of November 2018. The lander is studying the interior of Mars using Martian seismology and near-surface heat flow. Canadian researchers are part of the science team and ``will model the electrical conductivity of Martian rock, determine its water content, and ultimately shed light on the planet's interior structure and earliest history. Together with atmospheric information gathered by NASA's MAVEN satellite, they will use InSight's seismic, magnetic, and environmental data to track daily variations in Mars's magnetic field'' \citep{csa_insight} to understand the martian ionosphere.  Canadian researchers will also contribute to investigations of when and how space weather signals reach the surface of the planet, an important area of study for future astronaut-based exploration of the martian surface.

The CSA also contributed the Alpha Particle X-ray Spectrometer (APXS) to the Mars Science Laboratory (MSL's)\footnote{\url{https://mars.nasa.gov/msl/home/}} mission to analyze the composition of Martian rocks and soil \citep{msl_science}. Curiosity, the MSL rover,  has been operating on the Martian surface since 2012, with Canadian involvement on the research teams.

Canadian scientists were directly involved in the discovery and New Horizons\footnote{\url{http://pluto.jhuapl.edu/}} flyby of 2014~MU$_{69}$, which resolved a small, primitive Kuiper Belt object for the first time \citep{newhorizons_2014mu69}.  Such objects never experienced substantial heating and preserve information from the Solar System's formation.  The surface can further be used to test models for the small body population in the Kuiper Belt. 

The MESSENGER\footnote{\url{https://solarsystem.nasa.gov/missions/messenger/in-depth/}} spacecraft entered orbit about Mercury in 2011, and during its mission, completed 100\% mapping of the planet's surface and studied the planet's magnetic field, geology, and surface composition.  A Canadian-led team used MESSENGER data to lead and contribute to a wide range of studies that have transformed our understanding of Mercury's internal field and magnetospheric environment summarized in \citep{johnson_mercury_book, korth_mercury_book}.  This team led the discovery of crustal magnetization, suggesting a strong dynamo in the Mercury's early history \citep{johnson_mercury}. 

Much existing Canadian involvement in space probe missions is fortuitous and temporary, relying on haphazard arrangements such as co-appointments at US institutions. Critically needed are ways to encourage further Canadian participation (see recommendations at end of paper).

\subsection{Meteorites}

Canadians currently have one of the best networks of fireball cameras and meteorite recoveries, allowing specific meteorites to have their orbits and dynamical histories reconstructed.  These meteorites, once recovered, can be chemically analyzed in a laboratory setting, giving a detailed window on the formation conditions and history of the solar system.


\subsection{Telescopic Observations of Small Bodies}

\subsubsection{Asteroids} \label{asteroids}

Asteroid discovery, tracking, and orbital determination in the last decade has primarily been done by ground-based surveys with a growing contribution from space-based observatories including the Canadian Space Agency's Near-Earth Object Surveillance Satellite (NEOSSat)\footnote{\url{http://www.asc-csa.gc.ca/eng/satellites/neossat/default.asp}}. Orbital distributions combined with dynamical simulations gives constraints on formation and dynamical evolution of solar system architecture, while ground-based spectroscopy gives composition/formation information.  

Though many asteroids are initially seen by large survey programs such as Pan-STARRS and the Catalina Sky Survey, CFHT has provided critical follow-up observations that allow a fuller characterization and the extraction of critical science the large surveys are simply not equipped to do.  The loss of CFHT in the near-future could be mitigated by full access to LSST \citep[e.g.][]{Jonesetal2018}, that is, not simply access to the public data products that will be produced by LSST but to internal 'pixel' data that will not be generally available (see recommendations at end of paper).  MSE could provide spectra and compositions for thousands of asteroids, significantly increasing understanding of the formation location and thermal evolution of many asteroid populations.  

\subsubsection{Kuiper Belt Objects} \label{tnos}

Thousands of Kuiper Belt Objects (also called trans-Neptunian objects; TNOs) have now been discovered.  The most successful large TNO survey ever (the Outer Solar System Origins Survey; OSSOS) was led by Canadian principle investigators on the Canada-France-Hawaii Telescope (CFHT) \citep{Bannisteretal2018}.  This set of surveys detected nearly 1/3 of all known TNOs, including over half of those with well-measured orbits.  Part of the success of this survey has been careful dynamical classifications of discovered TNOs, primarily performed on Canadian computational facilities.  

CFHT's wide-field Megacam imager has been instrumental in discovering a large number of TNOs.  Once CFHT transitions to MSE (see section~\ref{mse}), Canada will no longer have access to a wide-field imager on a moderately sized telescope.  LSST is going to discover thousands of new TNOs, which will be useful for testing the details of giant planet migration.  Current large dynamical simulations are testing whether OSSOS observational data can lend support to or rule out different migration timescales, aand if it can place limits on the number of large planetesimals that were present at the time of Neptune's migration \citep[e.g.][]{Lawleretal2019}.

Even with the large number of TNOs expected to be discovered by LSST, deeper ground-based surveys will be necessary to detect the most distant TNOs.  Detecting these most distant TNOs will be necessary to answer questions about the structure of the outer solar system, such as whether there is an additional undiscovered giant planet \citep{p9}, or whether it is merely the result of observational biases \citep{kavelaarsp9}.  

The CADC has provided support for ground based TNO surveys by hosting data and data processing.  Testing measured orbital structure against predictions from dynamical models requires extensive dynamical simulations.  TNO modelling has made use of computational resources provided by CANFAR \citep[e.g.][]{pl17}.  CADC and CANFAR are both fantastically useful for TNO science and need continued support from the astronomy community.

\subsection{Space Weather}

Space weather can be defined as time-varying conditions driven by disturbances in the solar wind that affect planetary magnetospheres, ionospheres, and exospheres. It affects all planets in our Solar System (less strongly outer solar system planets), and likely affects exoplanets especially around magnetically active stars. Most space weather is driven by coronal mass ejections (CMEs) from the Sun, which here at Earth can cause geomagnetic storms. Such storms have the potential to significantly disrupt spacecraft electronics, communication and navigation systems, power grids, and even pipelines. Given humanity's reliance on technology, the past few decades have seen a large effort towards better understanding space weather through space-based measurements, ground and space-based monitoring of the Sun, and modeling. 

One of the ``holy grails" of the space weather community is to perfect geomagnetic storm prediction and build an early warning system for geomagnetic storms, thereby protecting our space and technology assets. This requires many more instruments in space than currently available. The Canadian space weather community has successfully used collaborations with the astronomy community to piggyback instruments on missions not dedicated for space weather to conduct highly influential research towards this goal. Cross-fertilization of ideas has recently enabled a soft X-ray space weather mission in which astronomers' noise becomes a space physics signal.

\section{Future Telescopic and Computational Facilities that will Benefit Canadian Planetary Science}

With the previous science topics in mind, we now discuss up-coming telescopic facilities. This discussion is not comprehensive, but rather focuses on two facilities that take focus of a substantive number of other white papers: the Large Synoptic Survey Telescope, and the MaunaKea Spectroscopic Explorer.

\subsection{Nearly all areas of Planetary Science will be Enhanced by Access to LSST}

The Large Synoptic Survey Telescope is an upcoming 8.4~m telescope being built on Cerro Pachon (near Gemini-South) with the singular purpose of executing a 10-year, 6-band (ugrizY) optical survey, with first light expected in 2022. One of the defining characteristics of this survey is frequent repeats to a given sky region. While the exact details of the survey patterns are still yet to be determined, the general survey strategy will involve repeat visits to a region in a given filter within four nights, with the next visit occurring roughly 10 days later. Over the full duration of the survey, the expected stack depth is roughly $r=28$. See LRP2020 papers by Hlozek and Fraser for further discussions on the LSST operations and expected outputs.

One of the four main science drivers for the LSST is to provide a detailed inventory of the Solar System's minor planet populations. Generally, the LSST will provide an order of magnitude increase in the overall inventory for all classes of planetesimal, including but not exclusively, asteroids, comets, and Kuiper Belt Objects. The LSST will probe to  a completeness magnitude that is roughly a full magnitude fainter in depth compared to past surveys for each of these populations, while simultaneously gathering high quality astrometry and photometry of each detected source, from which data products will be extracted including high quality orbits, lightcurves, surface colours, just to name a few. Critically, the LSST will survey the entire Solar System below $\sim30^\circ$ ecliptic latitude, in such a way that detection biases relating to object type, orbit, and class, can be reliably extracted.

In short, the LSST will address all shortcomings of past wide field moving object surveys, and immediately provide data products of a quality that is rarely provided from those surveys. Given the order of magnitude increase in number of detections, and the unprecidented data quality the survey will provide, it is fair to say that the LSST will revolutionize nearly all aspects of observational astronomy of the Solar System.

Canada has had a long and successful history in ground-based wide-field astronomy. In particular, Canada remains invested in wide-field searches for moving bodies (see Sections~\ref{asteroids} and \ref{tnos}). Since its inception, the Canada-France-Hawaii Telescope has been the Canadian community's main wide-field survey instrument. Currently, MegaCam on the CFHT provides Canada's only direct access to a wife-field imaging facility. Having just passed its 40th anniversary of CFHT's first light, and without any optical imager updates in years, the CFHT is showing its age, and is now being highly outperformed by other survey instruments such as the Dark Energy Camera on the Blanco telescope at the Cerro-Tololo Inter-American Observatory . 

Due to its age, and the natural evolution of Canadian astronomy interests, the CFHT is due to be shuttered, and will likely be replaced by the MaunaKea Spectroscopic Explorer. Without investment in appropriate facilities, Canada will soon be without any wide-field imaging facility, a situation that does not reflect Canadian scientific interests. Fortunately, a scenario has been put forth to rectify this situation. The concept of a Canadian hosted, LSST-light data centre is presented in the white paper by Fraser et al. The so-called LSST-light datacentre would host the publicly released data products most desired by the Canadian astronomy community, including the transient catalogs, which will contain photometric and astrometric data of the moving sources detected by the LSST. If funded, the LSST-light datacentre would satisfy many, if not most, of the nation's requirements for classical optical wide-field astronomy, at a fraction of the cost for a national buy-in. For the planetary science community in particular, the data products hosted at the LSST-light data centre, combined with the orbital catalogs that will be publicly available at the Minor Planet Centre, will enable Canadians to study virtually all aspects of planetary science that would be made possible for direct LSST members. 

\subsection{Canadian Computational Facilities are Vital}

Recently, the LSST Corporation adjusted the international membership model from a direct 
cost per membership, to  in-kind contributions. To that end, possible contributions from the Canadian community have been sought. One possible contribution under discussion is for CADC to host the public data set for world wide distribution.  This contribution would also provide Canadian astronomers with what has is referred to as an  LSST-light data centre (see the LSST Data Access white paper). 
The LSST-light data centre storage requirement is 18 PB with significant computing capacity to delivery the significant database content contained within those 18PB. 
To provide some perspective, the current storage capacity\footnote{the capacity of the archive is about 60\% of the aw storage capacity due to erasure encoding of the data} of CADC is approximately $\sim2$~PB and expected to grow to approximately 5~PB over the next 3 years.
For comparison, the Canadian SKA Regional Centre is anticipated to require a capacity of 42~PB (not counting near-line capacity). 

To enable the LSST-light data centre to provide useful function to the solar system community it is essential that pixel data be easily accessible.  This direct access will enable determination of the physical characteristics of the detected objects (e.g. binarity, coma, confusion with background etc.) and allow detailed science investigation. 
This can be achieved by having the storage for the LSST-light datacentre co-located with processing provided to the Canadian LSST Alert Science Platform (CLASP).  CLASP will act as a science portal to the LSST content while the CADC will manage the imaging and catalog collection.  
As the LSST light data centre will be collocated within a national research data centre, the costs for hosting this data will be incremental and the management of that data will be incremental to the activities of the CADC, thus minimizing costs.

\subsection{MSE Spectroscopy for Solar System Bodies} \label{mse}
The MaunaKea Spectroscopic Explorer\footnote{\url{https://mse.cfht.hawaii.edu/}} is a concept telescope, that would see the CFHT replaced with an 11-m, highly multi-plexed, fibre-fed UV-optical-NIR spectrograph. For every exposure, MSE will gather  spectra of up to 4,322 sources in a given 1.5 square degree area. MSE will act as a national observatory facility for use by all Canadians and international partners. For more details on MSE, see the LRP2020 paper by Hall et al.

In many ways, the MSE is for spectrographs what the LSST is for wide-field imaging facilities. With a larger aperture, large areal coverage, and very high multiplex capabilities, MSE reflects a massive increase in spectroscopic etendu. Though admittedly, moving objects present a particular challenge. This is primarily as a result of the instrument's inability to simultaneously track at the different rates of motion exhibited by individual moving targets in a field. MSE will be able to track at a single rate of motion, be that for a single minor body, or sidereal. Admittedly, the vast majority of exposures will be gathered with the latter, and so, the lack of differential tracking will effectively limit the exposure times for a given source to the time it takes to traverse roughly half a fibre's apparent width. 

MSE fibres are likely to come in two sizes: 0.8'' for the high resolution spectrograph, and 1'' for the low resolution spectrograph. Minor planet rates of motion are highly variable, and depend on the distance to a source, and its orbital geometry, but for example, at opposition, an asteroid will exhibit an apparent rate of motion of $\sim45"/hr$ while a Kuiper Belt Object will exhibit a much slower $\sim4"/hr$. For both of these targets, exposure times are limited to 1.3 and 15 minutes, respectively\footnote{The exposure time could be extended by enabling fibres to move within their patrol fields at a few arc-secconds per hour. This capability is not included in the preliminary design of the MSE fibre positioner system but is technically achievable given the current design.}. Thus, at a useful SNR level, MSE spectra will be limited to asteroids with brightnesses $g\lesssim22$, and Kuiper Belt objects with $g\lesssim24$. Even at these depths however, we expect nearly 10,000 KBOs to be within reach -- more than twice the currently known number -- and hundreds of thousands of asteroids, not to mention other bodies like comets, Trojans, etc. Like the LSST, the MSE will present a fundamental step forward in the spectral science of small bodies. 

\subsection{A Canadian Microsatellite for TNO Occultations}
A Canadian Micro-satellite mission with a small 15-cm optical telescope payload can make significant contributions to planetary science.  A payload capable of short, sub-second exposures can obtain target-of-opportunity observations of TNO stellar occultations.  In the coming Large Synoptic Survey Telescope (LSST) era, the number of known sources will increase by an order of magnitude or more, but our capacity to determine the physical sizes of these objects continues to lag.  With the recent availability of precise stellar astrometry from the European Gaia satellite, it is now possible to tie TNO orbits to a stellar reference frame whose precision is on the same angular scale as the projected sizes of TNOs. We are thus entering an era where accurate predictions of stellar occultation events are becoming possible. A small-sat telescope could provide a critical platform for observing these events and measuring TNO sizes. From a low altiude (400-100 km) polar sun-synchronous a Canadian microsat mission can obtain high-duty cycle continuous rapid cadence observations of predicted occultations.  These goals merge well with requirements of the proposed Photometry Observations of Extrasolar Planets (POEP) Microsatellite mission.  The POEP mission is a microsatellite with a 15-cm telescope that feeds dual frame-transfer devices to obtain high-precision and duty cycle u and i-band photometry.  Based the heritage of MOST \citep{Walker2003} and NEOSSat \citep{Laurin2008} the POEP will provide 25\% Guest Observer time to the Canadian community to support research in exoplanets, stellar and planetary astrophysics.  

\section{Summary and Recommendations}

Here we summarize our recommendations for what to support in the next decade for the best Canadian planetary science possible:

\begin{itemize}
    \item \textbf{Join LSST}: LSST will enhance nearly all areas of planetary science in Canada by providing good orbits for NEOs, asteroids, TNOs, and comets.  This will provide data both for new fundamental studies of the formation and evolution of our solar system, and also orbital data support for space missions.  
    \item \textbf{Support CADC to host LSST data}: Providing a data centre to host LSST data can be Canada's in-kind contribution to join. The costs for hosting and managing the data will be incremental to the current activities of CADC, and will enrich CADC and CANFAR's current role as a computational hub for orbital measurements in the outer Solar System.
    \item With consent from Native Hawaiians (see LRP paper on Canadian Astronomy on Maunakea), we support the conversion of CFHT into MSE, because it will allow spectroscopy on many thousands of small bodies, leading to understanding of the composition and evolution of these asteroids, comets, and TNOs.
    \item We support the proposed Canadian-built microsatellite POEP, which will provide occultation measurements of TNOs, measuring the size distribution and answering questions on formation and collisional evolution.
    \item  We recommend that there be a process in place for astronomers and space physicists to coordinate planning planetary missions, as they may be able to piggyback science instruments that are highly consequential for space weather studies on such missions.
    \item We recommend supporting the participation of Canadian researchers in international planetary and planetary astronomy missions, e.g. by providing support for competitively-selected researchers through NASA’s Participating Scientist and ESA’s Guest Investigator programs.  Upcoming  opportunities include future Discovery and New Frontiers Missions as well as the Bepi-Colombo mission.
\end{itemize}



\begin{lrptextbox}[How does the proposed initiative result in fundamental or transformational advances in our understanding of the Universe?]
Our top initiative, LSST, will discover thousands of new small bodies in the Solar System, providing new constraints on the formation and evolution of our Solar System.

\end{lrptextbox}

\begin{lrptextbox}[Are the associated scientific risks understood and acceptable?]
N/A

\end{lrptextbox}

\begin{lrptextbox}[Is there the expectation of and capacity for Canadian scientific, technical or strategic leadership?] 
Canadian computational facilities, CADC and CANFAR have hosted and processed data from observatories all over the world.  Canada is well-poised to host LSST data for Canadian astronomers to access.

MOST and NEOSSat have established Canadian scientifitic, technical and strategic leadership in the use of small satellites for planetary science.

\end{lrptextbox}

\begin{lrptextbox}[Is there support from, involvement from, and coordination within the relevant Canadian community and more broadly?] 
Roughly 30\% of Canadian astronomers desire access to at least some aspects of LSST data products  (see LRP2020 papers by Hlozek and Fraser), all of which support the concept of the LSST-light data centre. Active involvement from both academic and NRC astronomers has helped rapidly evolve this idea.

Observations by a small satellite, especially occultations, have very similar requirements to exoplanet discovery and characterization, which several astronomers in Canada are already actively involved in (see exoplanet LRP2020 paper).

\end{lrptextbox}

\begin{lrptextbox}[Will this program position Canadian astronomy for future opportunities and returns in 2020-2030 or beyond 2030?] 
Starting in 2022, LSST will discover thousands of new objects that will require follow-up observations to fully characterize.  Orbital distributions will need to be modelled, and observation biases will need to be modelled to understand true, unbiased distributions.  There is a lot of computational modelling to be done by Canadian astronomers on facilities like CANFAR.

POEP is striving for a mission launch of 2025.  TNO studies could be an important scientific component of the mission.  

\end{lrptextbox}

\begin{lrptextbox}[Is the cost-benefit ratio, including existing investments and future operating costs, favourable?] 
The proposed LSST-light data centre at a minimum requires significant upgrade to Canadian astrophysical data storage facility. Such an upgrade will take advantage of the common services approach to keep costs to a minimum, and will likely leverage the vast experience of the CADC. 

 Canadian microsatellite missions leverage off existing infrastructure investments and capabilities.  This makes future operating costs extremely favourable. These types of missions can be accomplished without significant new investment from government.  

\end{lrptextbox}

\begin{lrptextbox}[Are the main programmatic risks
understood and acceptable?] 
The main constraints appear to be computational, but incrementally increasing an existing data centre (CADC) will minimize any risks.

\end{lrptextbox}

\begin{lrptextbox}[Does the proposed initiative offer specific tangible benefits to Canadians, including but not limited to interdisciplinary research, industry opportunities, HQP training,
EDI,
outreach or education?] 
The LSST-light data centre initiative  will likely lead to considerable in-kind contribution by Canadians to the LSST project, and thus directly facilitate membership and data access rights to the LSST for some Canadians.

LSST data hosting and processing will require training researchers and computer scientists.  Opportunities to use machine-learning algorithms and other cutting-edge Big Data techniques will provide Canadian students with highly employable software/data science skills.  

 New missions create good, well paying middle-class jobs.  Canadian-led microsat Missions require HQP and HQP training and are able to retain HQP in Canada.  Missions require Canadian investment in Canadian missions, and a Canadian-led mission should be a high priority for the entire astronomy community.  The planetary community will scientifically benefit from access to such a facility.  New missions create capacity which enable reuse of facilities to further push techology and scientific discovery.  

\end{lrptextbox}


\end{document}